\documentclass[twoside,twocolumn,9pt]{article}
\usepackage{extsizes}
\usepackage[super,sort&compress,comma]{natbib} 
\usepackage[version=3]{mhchem}
\usepackage[left=1.5cm, right=1.5cm, top=1.785cm, bottom=2.0cm]{geometry}
\usepackage{balance}
\usepackage{widetext}
\usepackage{times}
\usepackage{sectsty}
\usepackage{graphicx} 
\usepackage{lastpage}
\usepackage[format=plain,justification=raggedright,singlelinecheck=false,font={stretch=1.125,small,sf},labelfont=bf,labelsep=space]{caption}
\usepackage{float}
\usepackage{fancyhdr}
\usepackage{fnpos}
\usepackage[english]{babel}
\usepackage{array}
\usepackage{droidsans}
\usepackage{charter}
\usepackage[T1]{fontenc}
\usepackage[usenames,dvipsnames]{xcolor}
\usepackage{setspace}
\usepackage[compact]{titlesec}

\usepackage{bm}
\usepackage{amsmath, amssymb}

\definecolor{cream}{RGB}{222,217,201}

\newcommand{\MA}{\mathsf{A}}
\newcommand{\MB}{\mathsf{B}}
\newcommand{\Deff}{D_{\textrm{eff}}}
\newcommand{\Dbare}{D_0}

\allowdisplaybreaks[1]

\begin{document}

\pagestyle{fancy}
\thispagestyle{plain}
\fancypagestyle{plain}{

}

\makeFNbottom
\makeatletter
\renewcommand\LARGE{\@setfontsize\LARGE{15pt}{17}}
\renewcommand\Large{\@setfontsize\Large{12pt}{14}}
\renewcommand\large{\@setfontsize\large{10pt}{12}}
\renewcommand\footnotesize{\@setfontsize\footnotesize{7pt}{10}}
\makeatother

\renewcommand{\thefootnote}{\fnsymbol{footnote}}
\renewcommand\footnoterule{\vspace*{1pt}%
\color{cream}\hrule width 3.5in height 0.4pt \color{black}\vspace*{5pt}} 
\setcounter{secnumdepth}{5}

\makeatletter 
\renewcommand\@biblabel[1]{#1}            
\renewcommand\@makefntext[1]%
{\noindent\makebox[0pt][r]{\@thefnmark\,}#1}
\makeatother 
\renewcommand{\figurename}{\small{Fig.}~}
\sectionfont{\sffamily\Large}
\subsectionfont{\normalsize}
\subsubsectionfont{\bf}
\setstretch{1.125} 
\setlength{\skip\footins}{0.8cm}
\setlength{\footnotesep}{0.25cm}
\setlength{\jot}{10pt}
\titlespacing*{\section}{0pt}{4pt}{4pt}
\titlespacing*{\subsection}{0pt}{15pt}{1pt}


\makeatletter 
\newlength{\figrulesep} 
\setlength{\figrulesep}{0.5\textfloatsep} 

\newcommand{\topfigrule}{\vspace*{-1pt}%
\noindent{\color{cream}\rule[-\figrulesep]{\columnwidth}{1.5pt}} }

\newcommand{\botfigrule}{\vspace*{-2pt}%
\noindent{\color{cream}\rule[\figrulesep]{\columnwidth}{1.5pt}} }

\newcommand{\dblfigrule}{\vspace*{-1pt}%
\noindent{\color{cream}\rule[-\figrulesep]{\textwidth}{1.5pt}} }

\makeatother


\twocolumn[
  \begin{@twocolumnfalse}
\sffamily

\noindent\LARGE{\textbf{Hydrodynamics of bilayer membranes with diffusing transmembrane proteins}} \\

\noindent\large{Andrew Callan-Jones, Marc Durand$^{\ast}$ and Jean-Baptiste Fournier} \\

\noindent\normalsize{We consider the hydrodynamics of lipid bilayers containing transmembrane proteins of arbitrary shape.  This biologically-motivated problem is relevant to the cell membrane, whose fluctuating dynamics play a key role in phenomena ranging from cell migration, intercellular transport, and cell communication.   
Using Onsager's variational principle, we derive the equations that govern the relaxation dynamics of the membrane shape, of the mass densities of the bilayer leaflets, and of the diffusing proteins' concentration. With our generic formalism, we obtain several results on membrane dynamics.  We find that proteins that span the bilayer increase the intermonolayer friction coefficient.  The renormalization, which can be significant, is in inverse proportion to the protein's mobility.  Second, we find that asymmetric proteins couple to the membrane curvature and to the difference in monolayer densities.  
For practically all accessible membrane tensions ($\sigma> 10^{-8}$ N/m) we show that the protein density is the slowest relaxing variable.  Furthermore, its relaxation rate decreases at small wavelengths due to the coupling to curvature.  We apply our formalism to the large-scale diffusion of a concentrated protein patch.  We find that the diffusion profile is not self-similar, owing to the wavevector dependence of the effective diffusion coefficient.} \\


\end{@twocolumnfalse} \vspace{0.6cm}

  ]

\renewcommand*\rmdefault{bch}\normalfont\upshape
\rmfamily
\section*{}
\vspace{-1cm}


\footnotetext{Universit\'e Paris Diderot, Sorbonne Paris Cit\'e, Laboratoire Mati\`ere et Syst\`emes Complexes (MSC), UMR 7057 CNRS, F-75205 Paris, France}

\date{\today}

\section{Introduction}
\label{Intro}

Biological membranes are lipid bilayers forming the envelopes of plasma membranes, nuclei, organelles, tubules and transport vesicles within a cell~\cite{lodish_book}. They are versatile structures, both fluid and elastic, that can change shape or topology in order to accomplish the cell functions. From the dynamical point of view, membranes can be viewed as a system of four fluid phases in contact: a pair of two-dimensional (2D) lipid phases and two three-dimensional (3D) aqueous phases. These phases, separated but strongly coupled~\cite{Evans94CPL,Seifert93EPL}, exhibit nontrivial multiphase flow behaviors~\cite{Seifert93EPL,Fournier09PRL,Brennen_book}.
The dynamics of bilayer membranes containing transmembrane proteins at a high concentration are especially challenging because the proteins form a fifth phase that effectively interdigitates two components of the multiphase flow. This situation corresponds to the actual biological one, where macromolecular crowding effects are ubiquitous and which are known to make molecules in cells behave in radically different ways than in artificial lipid vesicles~\cite{lodish_book}. 

The study of the diffusive behaviour of proteins embedded in a membrane, taking into account the hydrodynamics of the membrane and that of the surrounding solvent, originated with the seminal work of Saffman and Delbr\"uck on cylindrical inclusions in flat membranes~\cite{Saffman:1975aa}.  Since then, investigations into the effects on single protein diffusion both of membrane height fluctuations and of the coupling between membrane curvature and protein shape have been carried out. While earlier work found that membrane fluctuations  tend to enhance the diffusion coefficient $D$ of curvature--inducing proteins~\cite{Reister:2005,Leitenberger:2008aa}, more recent studies that take greater account of the surrounding membrane deformation caused by a protein, showed that $D$ is actually reduced~\cite{Naji:2009aa,Reister-Gottfried:2010aa}. This predicted lowering of $D$ has been recently verified experimentally~\cite{Quemeneur14}.

Despite these advances, there has been relatively little work done on the collective diffusive behavior of many, interacting transmembrane proteins, taking into account the bilayer structure of the membrane and in turn the influence of the proteins on the in-plane and out-of-plane membrane dynamics.  In fact, almost thirty years ago, it was proposed that the energetic coupling between a curved membrane and asymmetric proteins would effectively result in greater attraction between proteins~\cite{Leibler:1986,Leibler:1987}, suggesting a reduction in the cooperative diffusion coefficient.
However, the membrane was treated as a single, mathematical surface, thus neglecting the influence of transmembrane proteins on the coupling between the two monolayers.  The influence of the bilayer structure of the membrane on single protein diffusion has recently been studied, in particular for proteins located in one of the two monolayers~\cite{Camley:2013}, yet no comparable work at the many protein level has been proposed.
We note that though a very general theoretical framework was developed some time ago~\cite{Lomholt:2005aa}, a handy theory describing the multiphase dynamics of a deformable membrane bilayer with diffusing transmembrane proteins is still lacking.
 
Simple, single-phase bilayer membranes already have a complex hydrodynamic behavior~\cite{Kramer71JCP,Brochard75JPhys} due to their soft out-of-plane elasticity~\cite{Helfrich73}. The complete equations describing the hydrodynamics of bilayer membranes, including curvature and in-plane elasticities, intermonolayer friction, monolayer 2D viscosity and solvent 3D viscosity, were first derived by Seifert and Langer for almost planar membranes~\cite{Seifert93EPL}. The method employed was a careful balance of in-plane and out-of-plane elastic and viscous stresses. Generalization to non-linear membrane deformations was achieved by Arroyo \textit{et al.}~\cite{Arroyo09PRE,Rahimi12PRE,Rahimi13SM} using covariant elasticity and Onsager's variational principle~\cite{Goldstein_book,Doi_Onsager}; see also Ref.~\cite{Fournier15IJNM}. Importantly, the monolayers must be treated as compressible fluids. Indeed, while the 3D density of the lipid region remains almost constant, the 2D lipid density can significantly change because the membrane thickness is free to adapt \cite{Statistical2003Safran}. Moreover, any lipid density difference between the two monolayers couples with the membrane curvature~\cite{Svetina85,Miao94PRE}. In the early studies of membrane hydrodynamics the bilayer structure of the membrane was neglected~\cite{Kramer71JCP,Brochard75JPhys}. While this is a good approximation~\cite{Fournier15IJNM} for tensionless membranes at length-scales much larger than microns, experiments and theoretical studies have shown that taking the bilayer structure into account is essential at shorter length-scales~\cite{Evans94CPL,Seifert93EPL,Sens04PRL,Fournier09PRL,Bitbol12SM}. This is chiefly due to the importance of the dissipation due to the \textit{intermonolayer friction},
caused by the relative motion of the lipid tails, occurring when the two monolayers have different velocities~\cite{Evans94CPL,Otter:2007aa}.

In this paper, we derive the coupled equations describing the multi-phase flow of a deformable bilayer membrane hosting transmembrane proteins that diffuse collectively. For a one-component, or even a two-component lipid bilayer membrane, the hydrodynamical equations for each monolayer can be written in a relatively simple manner by balancing standard 2D and 3D hydrodynamic stresses with the in-plane pressure gradient and the intermonolayer stress~\cite{Seifert93EPL,Okamoto15condmat}. This is no longer possible for a bilayer membrane with spanning proteins. Indeed, the two monolayers are not only coupled through the intermonolayer stress, they are also coupled through the protein mass conservation law, manifested by the equality of the protein fluid velocity in the two monolayers, which connects the diffusive flows of both monolayers. This constraint produces additional in-plane stresses that cannot be understood simply. Using Onsager's variational principle, because it is based on minimizing the total energy dissipation under all the relevant constraints, we are able to determine the constitutive equations consistently and to uncover these nontrivial stresses.

One simple consequence, that is readily understood, is that with respect to the hypothetical situation where the proteins would be broken into independent halves living in each monolayer, the intermonolayer friction coefficient $b$ increases in inverse proportion to the proteins' mobility $\Gamma$. Indeed, imagine the simple case of up-down symmetric proteins within a bilayer flowing in such a way that the two monolayers have exactly opposite velocities.  By symmetry, the proteins remain immobile, and friction arises not only from the lipid tails of the contacting monolayers, but also from the dissipation ($\propto\! 1/\Gamma$) due to the flow within the monolayers between the lipids and the proteins.

From our coupled dynamical equations, we obtain the collective diffusion coefficient $D_\mathrm{eff}$ of up-down asymmetric proteins and its wavevector dependence. Relaxation of small wavevector ($q$) disturbances of the density of asymmetric transmembrane proteins couples to the relative motion of the monolayers, which involves intermonolayer friction.  This leads to a $b$--proportional reduction of the effective diffusion coefficient $D_\mathrm{eff}(q)$, as compared with equivalent but symmetric proteins.  In contrast, for large wavevectors, the energetic coupling between protein density and membrane curvature further reduces the effective diffusion coefficient $\Deff(q)$, in agreement with Refs.~\cite{Leibler:1986,Leibler:1987}.  The crossover between the two regimes occurs at wavevector $q_{\textrm{c}}\sim \sqrt{\sigma/\kappa}$, with $\sigma$ the membrane tension and $\kappa$ the membrane bending modulus. Below $q_{\textrm{c}}$, tension effectively flattens the membrane, thus only the protein concentration and the difference in monolayer densities couple; above $q_{\textrm{c}}$, 
short scale membrane deformations favor spatially inhomogeneous protein concentration, thus lowering $\Deff$.  Finally, we confirm the wavevector dependence of the protein diffusion coefficient by examining, with our formalism, the dynamics of an initially localized excess concentration of proteins.

The present article is organized as follows. In Sec.~\ref{Hydro_description}, using Onsager's variational principle~\cite{Doi_Onsager}, we derive the general equations describing the multiphase flow of the system. In Sec.~\ref{Relax_modes}, we calculate the relaxation modes of the dynamical equations coupling the membrane shape, the lipid density--difference and the protein density. We discuss the role of membrane tension and protein asymmetry, and we derive the wavevector--dependent effective diffusion coefficient of the proteins. In Sec.~\ref{Diffusion_spot}, we analyse the spreading of a concentrated spot of proteins and we discuss its anomalous diffusion. In Sec.~\ref{Conclusions}, we give our conclusions and discuss the perspectives of our approach.

\section{Hydrodynamic description}
\label{Hydro_description}

\begin{figure}
\centerline{\includegraphics[width=.9\columnwidth]{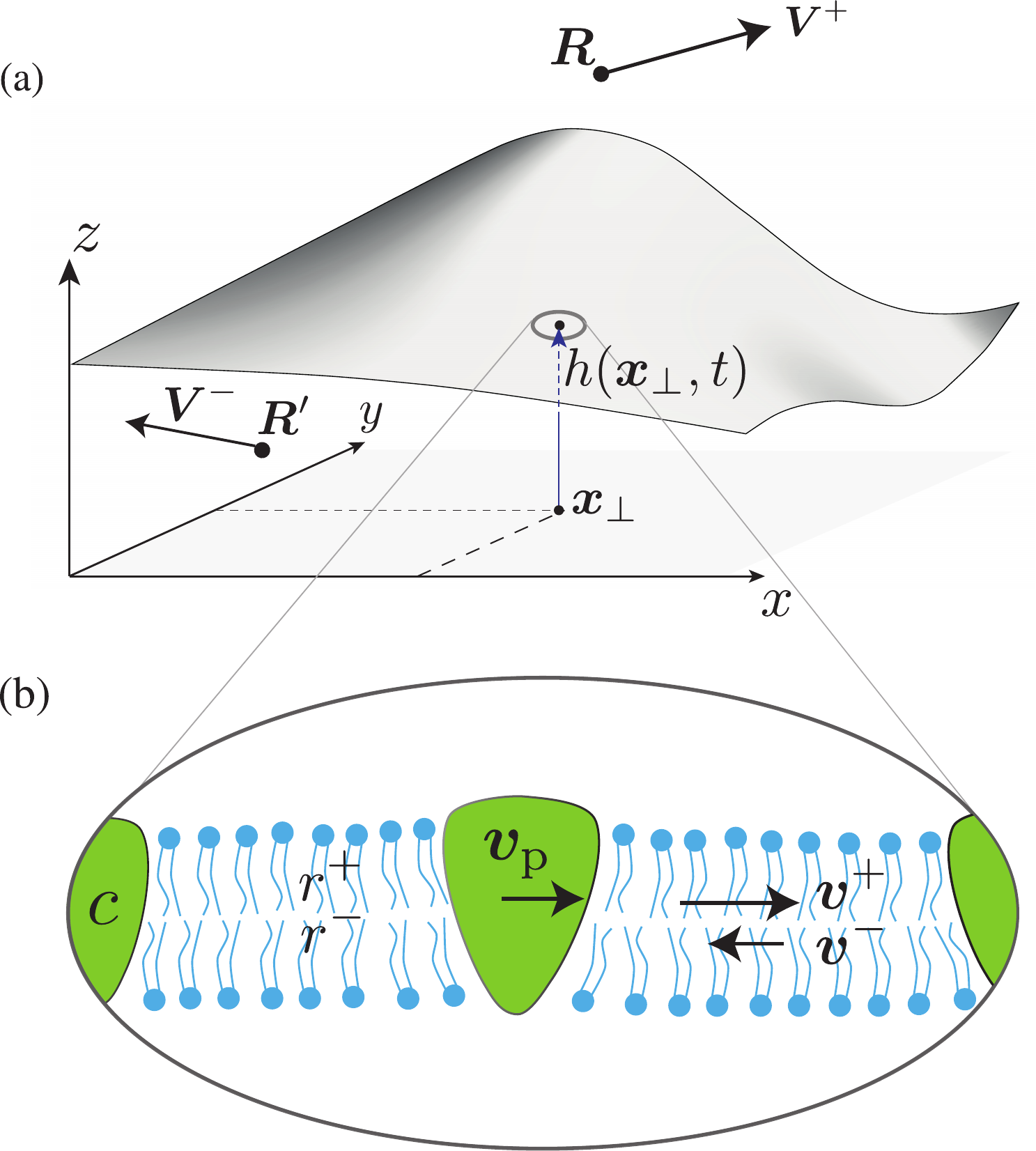}}
\caption{\label{figs:sketch}(color online.) Membrane with asymmetric, spanning proteins.  (a) Geometric description of a membrane with height $h$ and solvent velocities $\bm{V}^{\pm}$.  (b) Monolayers have total mass densities $r^{\pm}$ and barycentric velocities $\bm{v}^{\pm}$.  Proteins have total mass fraction $c$ and velocity $\bm v_{\textrm{p}}$. }
\end{figure}

To understand the coupling between the protein diffusion in a bilayer membrane and the membrane curvature, we first establish the dynamical equations governing the relaxation of small fluctuations in membrane shape, and densities of lipid and protein relative to a flat, homogeneous configuration.
Let us consider a one lipid species bilayer membrane hosting a non-uniform mass fraction $c$ of (identical) asymmetric transmembrane proteins. The space is parametrized in Cartesian coordinates by $\bm R=(\bm x_{\perp},z)$, with $\bm x_{\perp}=(x,y)$. The membrane shape is described, in the Monge gauge, by the height $h(\bm x_{\perp},t)$ of the bilayer's midsurface $\mathcal{M}$ above the reference plane $(x,y)$,  with $t$ the time. Quantities with the superscript $+$ and $-$ will refer to the upper and lower monolayer, respectively (Fig.~\ref{figs:sketch}). Considering one protein, we call $1/s^+$ (resp.\ $1/s^-$) the fraction of its mass lying in the upper (resp.\ lower) monolayer, with
\begin{align}
1/s^+ + 1/s^- = 1.
\end{align}
Let $r^\pm$ be the total (lipid+protein) mass density within each monolayer and $c^\pm$ the protein mass fraction within each monolayer. Then, $r^\pm c^\pm$ is the protein mass density in each monolayer, and $c^\pm$ is thus related to the protein mass fraction $c$ through
\begin{align}
s^\pm r^\pm c^\pm=(r^+ + r^-)c\,.
\label{massprot}
\end{align}
Note that $r^\pm$, $c^\pm$ and $c$ vary along the membrane, while $s^\pm$ are constant coefficients. Like in Ref.~\cite{Seifert93EPL}, all densities are defined on~$\mathcal{M}$.

We consider small deviations with respect to a flat membrane ground state with uniform protein mass fraction $c_0$ and mass densities $r_0^\pm$. 
Thus, we write
\begin{align}
&r^\pm=r_0^\pm(1+\rho^\pm),\\
&c^\pm=c_0^\pm+\phi^\pm,\\
&c=c_0+\phi.
\end{align}
The variables $h(\bm x_{\perp},t)$, $\rho^\pm(\bm x_{\perp},t)$ and $\phi(\bm x_{\perp},t)$ are our main variables and they will be considered as first-order quantities. Note that $c_0^\pm$, as well as $\phi^\pm$, follow from Eq.~(\ref{massprot}): 
\begin{align}
c_0^{\pm}=\frac{r_0^{+}+r_0^{-}}{r_0^{\pm} s^{\pm}} c_0\,,
\end{align}
and, to linear order in the densities,
\begin{align}
\phi^{\pm}&=\mp\frac{r_0^{\mp} c_0}{r_0^{\pm} s^{\pm}} \left(\rho
   ^+-\rho ^-\right)+\frac{r_0^++r_0^-}{r_0^{\pm} s^{\pm}} \phi\,+\mathcal{O}(2).
\label{eq:MonoProtDensITOBilayerDens}
\end{align}
That is, the monolayer protein mass fractions couple not only to $\phi$, but also to the difference in monolayer densities $\rho^+-\rho^-$.

Since the monolayers have little interactions apart from being pushed into contact by hydrophobic forces, the Hamiltonian of the bilayer can be written as
\begin{align}
\mathcal{H}&=
\int_\mathcal{M}\!dS\left[
g^+(H,r^+,c^+) + g^-(H,r^-,c^-) \right]
\label{H1}\\
&\simeq\int\! d^2x_{\perp}\left[
\frac{\sigma}{2}\left(\nabla h\right)^2
+\sum_{\epsilon=\pm} f^\epsilon(\nabla^2h,\rho^\epsilon, \phi^\epsilon)
\right],\label{H2}
\end{align}
where $dS$ is the elementary area and $H\simeq\nabla^2h/2$ the mean curvature of $\mathcal{M}$. Eq.~(\ref{H2}) is obtained by expanding Eq.~(\ref{H1}) at quadratic order. The membrane tension, $\sigma$, arises from the zeroth-order terms in $g^\pm$ mutiplied by $dS/d^2x_\perp\simeq1+(\nabla h)^2/2$. The most general quadratic form of $f^\pm$ can be written as~\cite{justif1}
\begin{align}
f^\pm&=\frac{\kappa^\pm}{4}\left(\nabla^2 h\right)^2
+\frac{k^\pm}{2}\left(\rho^\pm \pm e^\pm\nabla^2h +\beta^\pm\phi^\pm\right)^2
\nonumber\\
&\pm\lambda^\pm k^\pm e^\pm(\nabla^2h)\,\phi^\pm
+\frac{\alpha^\pm k^\pm}{2}\,{\phi^\pm}^2,
\label{fpm}
\end{align}
which generalizes the form proposed by Seifert and Langer for protein-free membranes~\cite{Seifert93EPL}. Note that, although the lipids are the same in both monolayers, the elastic constants $\kappa^\pm$, $k^\pm$, $e^\pm$, etc., are monolayer dependent if the proteins are up-down asymmetric. The Gaussian curvature elasticity term can be neglected thanks to the Gauss-Bonnet theorem despite the inhomogenity of the corresponding stiffness, $\bar\kappa^\pm$, arising from the proteins \cite{Brannigan:2007aa,West2009101}. Indeed, taking into account the dependence of $\bar\kappa^\pm$ in $\phi^\pm$ would result in a higher-order, cubic term, since the Gaussian curvature is already a second-order quantity. We shall assume throughout that the system is far from protein phase separation. Here $\kappa^\pm$ is the monolayer bending rigidity~\cite{Helfrich73}, $k^\pm$ the stretching modulus, $e^\pm$ a density--curvature coupling constant that can be interpreted as the distance between $\mathcal{M}$ and the monolayer neutral surface, and $\beta^\pm$, $\lambda^\pm$ and $\alpha^\pm$ are dimensionless coupling constants arising from the presence of the proteins, normalized by $k^\pm$. All these constants depend in principle on the background protein mass fraction $c_0$ and on the bare mass densities $r_0^\pm$, hence on the tension $\sigma$~\cite{Bitbol2012,Watson2013}, although these dependences should be moderate~\cite{note_moderate}. Identifying them would require a specific microscopic model, which is beyond the scope of this work.

Let us call $\bm V^\pm$ the solvent velocity on both sides of the membrane, $\bm v^\pm$ the in-plane barycentric velocity of the lipid+protein binary fluid, in each monolayer, and $\bm v_{\textrm{p}}^+=\bm v_{\textrm{p}}^-$ the barycentric velocity of the transmembrane proteins. These quantities are obtained from the total mass flux $\bm g^\pm$ and the protein mass flux $\bm g_{\textrm{p}}^\pm$ using the relations $\bm g^\pm=r^\pm \bm v^\pm$ and $\bm g_{\textrm{p}}^\pm=r^\pm c^\pm \bm v_{\textrm{p}}^\pm$. Our model is therefore valid only on length scales greater than the inter-protein spacing. We assume that the velocities are generated by the relaxation of the membrane only, hence they are also small quantities that can be considered parallel to $(x,y)$ at first-order. In the following, we use the convention that Latin indices represent either $x$ or $y$ while Greek indices represent either $x$, $y$ or $z$. Assuming no slip at the membrane surface and no permeation, we have on the membrane, at $z=0$:
\begin{align}
V_i^\pm=v_i^\pm\,,\label{Vivi}\quad
V_z=\dot h\,,
\end{align}
where a dot indicates partial time derivative. The mass continuity equations read~\cite{Landau_Fluid_Mechanics_book}:
\begin{align}
&\dot r^\pm+\partial_i(r^\pm v_i^\pm)=0\,,
\label{CoMContinuity}
\\
&r^\pm \dot c^\pm + r^\pm v_i^\pm \partial_i c^\pm + \partial_i J_i^\pm = 0\,,
\label{ProteinContinuity}
\end{align}
where the protein diffusion currents are defined by $\bm J^\pm=r^\pm c^\pm (\bm v_{\textrm{p}}^\pm-\bm v^\pm)$ and the last equation follows from
$\partial_t(r^\pm c^\pm)+\partial_i g^\pm_{p,i}=0$. We note that, with the help of Eqs.~(\ref{massprot}),~(\ref{CoMContinuity}), and \eqref{ProteinContinuity}, the protein continuity equation can be rewritten in terms of the bilayer protein mass fraction:
\begin{align}
r\dot{c}+r v_i\partial_i c&=-\frac{1}{2}\partial_i\left(s^+ J_i^++s^- J_i^-\right)+\frac{1}{2}\partial_i\left(c \Delta r\Delta v_i\right)\,,
\end{align}
where $r=r^+ + r^-$ is the total mass density, $v_i=(r^+ v_i^+ + r^- v_i^-)/r$ is the barycentric velocity, $\Delta r= r^+ - r^-$ and $\Delta v_i= v_i^+ - v_i^-$. Thus, the total protein density is convected by the bilayer barycentric velocity, as expected. A more striking observation is that the protein mass current relative to $v_i$ depends not only on the diffusive fluxes, but also on the relative monolayer velocity $\Delta v_i$.

A crucial element of this problem is the following. Since the proteins span the bilayer, $\bm v_{\textrm{p}}^+=\bm v_{\textrm{p}}^-$, and the monolayer protein currents must obey the following constraint:
\begin{align}
&s^+J_i^+ -s^-J_i^-
= (r^+ + r^-) c \,(v_i^- - v_i^+),
\label{Jconstraint}
\end{align}
as a consequence of Eq.~(\ref{massprot}) and the definition of the currents.
As we show below, this relation implies that transmembrane proteins slow down the relative motion between monolayers.  

In the Stokes approximation, i.e., neglecting all inertial effects, the dynamical equations of the whole system can be obtained by minimizing the total Rayleighian of the system, $\mathcal{R}=\frac{1}{2}(P_{\textrm{s}}+P_{\textrm{m}})+\dot{\mathcal{H}}$, equal to half the total dissipated power plus the time derivative of the Hamiltonian~\cite{Goldstein_book,Doi_Onsager}.
The dissipation in the solvent reads
\begin{align}
P_{\textrm{s}}=
\int_{z>0}\!\!d^3R\, \eta D_{\alpha\beta}^+D_{\alpha\beta}^+
+
\int_{z<0}\!\!d^3R\, \eta D_{\alpha\beta}^-D_{\alpha\beta}^-\,,
\end{align}
where $D_{\alpha\beta}^\pm=\frac{1}{2}(\partial_\alpha V_\beta^\pm+\partial_\beta V_\alpha^\pm)$. The dissipation within the membrane reads
\begin{align}
P_{\textrm{m}}&=\sum_{\epsilon=\pm}\int\!d^2 x_{\perp}
\left[
\eta_2 d_{ij}^\epsilon d_{ij}^\epsilon
+\frac{\lambda_2}{2}\,d_{ii}^\epsilon d_{jj}^\epsilon
+\frac{1}{2\Gamma}J_i^\epsilon J_i^\epsilon
\right]\nonumber\\
&+\int\!d^2 x_{\perp}\,\frac{b}{2}\left(v_i^+-v_i^-\right)^2\,,
\label{eq:DissipationMembrane}
\end{align}
where $d_{ij}^\pm=\frac{1}{2}(\partial_i v_j^\pm+\partial_j v_i^\pm)$, $\eta_2$ and $\lambda_2$ are the in-plane monolayer viscosities (compressible fluid), $\Gamma$ is the coefficient associated with the dissipation due to the diffusion currents~\cite{Landau_Fluid_Mechanics_book}, and $b$ is the intermonolayer friction coefficient~\cite{Evans94CPL,Seifert93EPL}. As in the case of the elastic constants, these dissipative coefficients depend on $c_0$ and $r_0^\pm$~\cite{Niemela,Oppenheimer:2009aa,Camley:2013,Camley2014}. Note that for the sake of simplicity we have assumed that $\eta_2$, $\lambda_2$, and $\Gamma$ are the same in both monolayers. The time derivative of the Hamiltonian is
\begin{align}
\dot{\mathcal{H}}=&\int\!d^2x_\perp
\Big\{\!
-\sigma\left(\nabla^2h\right)\dot h
\nonumber\\&~~~~~
+\sum_{\epsilon=\pm}\left[
f^\epsilon_\rho\dot\rho^\pm
+f^\epsilon_\phi\dot\phi^\pm
+\left(\nabla^2f^\epsilon_h\right)\dot h
\right]
\Big\},
\end{align}
where $f_h^\pm=\partial f^\pm/\partial(\nabla^2h)$, $f^\pm_\rho=\partial f^\pm/\partial\rho^\pm$, and $f^\pm_\phi=\partial f^\pm/\partial\phi^\pm$,

The Rayleighan $\mathcal{R}$ must be extremalized with respect to all the variables expressing the rate of change of the system: $\dot h$, $v^\pm_i$, $V^\pm_z$, $V^\pm_i$, $\dot\rho^\pm$, $\dot\phi^\pm$, $J^\pm_i$~\cite{Goldstein_book,Doi_Onsager}, and Lagrange multiplier fields must be introduced in order to implement the various constraints. Let $\omega(\bm x_\perp)$ be the Lagrange field associated with (\ref{massprot}), $\{\mu(\bm x_\perp),\gamma(\bm x_\perp),\zeta(\bm x_\perp),\chi(\bm x_\perp)\}$ associated with (\ref{Vivi}--\ref{ProteinContinuity}), and $\omega_i(\bm x_\perp)$ associated with (\ref{Jconstraint}). The bulk incompressibility condition $\partial_\alpha V_\alpha^\pm=0$ will be implemented by the Lagrange fields $p^\pm(\bm R)$ corresponding to the solvent's pressure. Performing the constrained extremalization of $\mathcal{R}$ yields the bulk Stokes equations:
\begin{align}
&-2\eta\partial_\beta D^\pm_{\alpha\beta}
+\partial_\alpha p^\pm=0\,,
\\
&\partial_\alpha V_\alpha^\pm=0\,.
\end{align}
and the surface equations in the plane $z=0$ of the unperturbed membrane (in the order of the dynamical variables listed above):
\begin{align}
&\nabla^2(f_h^+ + f_h^-)-\sigma\nabla^2h
+\gamma^++\gamma^-=0\,,
\\
&-2\eta_2\partial_jd_{ij}^\pm-\lambda_2\partial_i d_{jj}^\pm
\pm b\left(v_i^+-v_i^-\right)
-r^\pm\partial_i\zeta^\pm
\nonumber\\&~~~~~~~~~~~~~
+r^\pm\chi^\pm\partial_ic^\pm
\pm\omega_i (r^++r^-)c
+\mu_i^\pm=0\,,
\\
&\pm p^\pm \mp2\eta D_{zz}^\pm -\gamma^\pm = 0\,,
\\
&\mp 2\eta D^\pm_{zi}-\mu^\pm_i=0\,,
\\
&
f_\rho^\pm + r_0^\pm\zeta^\pm=0\,,\label{grototo}
\\
&
f_\phi^\pm
+r^\pm\chi^\pm=0\,,\label{grotata}
\\
&\Gamma^{-1}J_i^\pm-\partial_i\chi^\pm
\pm s^\pm \omega_i=0\,,
\label{grotiti}
\\
&\dot r^\pm+\partial_i\left(r^\pm v_i^\pm\right) = 0\,,
\\
&
r^\pm \dot c^\pm + r^ \pm v^\pm \partial_i c^\pm 
+\partial_i J_i^\pm = 0\,,
\\
&v_i^\pm-V_i^\pm=0\,,
\\
&\dot h-V_z^\pm=0\,,
\\
&s^+r^+c^+=s^-r^-c^-=(r^++r^-)c,\\
&s^+J_i^+ -s^-J_i^-
= (r^+ + r^-) c \,(v_i^- - v_i^+).
\end{align}
Note that in addition to $h$, $\rho^\pm$ and $\phi^\pm$, the velocities, the diffusion currents and, therefore, all the Lagrange multipliers are first-order quantities. Linearizing these surface equations, eliminating the Lagrange multipliers, the diffusion currents and $\phi^\pm$ (thanks to Eq.~(\ref{eq:MonoProtDensITOBilayerDens})) is straightforward and yields the following reduced set of surface equations:
\begin{align}
&\nabla^2f_h-\sigma\nabla^2h
+ p^+ - p^- -2\eta (D_{zz}^+ - D_{zz}^-)  = 0\,,
\label{eq1}\\
&-2\eta_2\partial_j d_{ij}^\pm -\lambda_2 \partial_i d_{jj}^\pm
\pm (b+\delta b)(v_i^+ - v_i^-)+\partial_i f_\rho^\pm
\nonumber\\&~~~~
\mp Q\left(s^+ r_0^- \partial_i f_\phi^+
- s^- r_0^+ \partial_i f_\phi^-\right)
\mp 2\eta D_{zi}^\pm = 0\,,
\label{eq2}\\
&\dot\rho^\pm+\partial_iv_i^\pm=0\,,
\label{eq3}\\
&\dot\phi-\Gamma S
\left(s^- r_0^- \nabla^2 f_\phi^+ + s^+ r_0^+ \nabla^2 f_\phi^-
\right)
\nonumber\\&~~~~
-T\partial_i(v_i^+ - v_i^-)=0\,,\label{eqdifus}
\end{align}
in which $f_h=f_h^++f_h^-$, $s^2=\frac{1}{2}[(s^+)^2 + (s^-)^2]\ge4$, $r_0=\frac{1}{2}(r_0^+ + r_0^-)$, $Q=c_0r_0/(s^2 r_0^+ r_0^-)$, $S=s^+ s^-/(4s^2 r_0^+ r_0^- r_0)$, $T=c_0[(s^+)^2r_0^+ - (s^-)^2r_0^-]/(4s^2 r_0)$, and 
\begin{align}
\delta b=\frac{2}{\Gamma}\left(\frac{r_0 c_0}{s}\right)^2\,.
\label{B}
\end{align}
These equations, Eqs.~(\ref{eq1}--\ref{eqdifus}), resemble the ones obtained by Seifert and Langer for a one-component lipid bilayer membrane~\cite{Seifert93EPL}, yet there are fundamental differences. These arise from the terms proportional to $\delta b$ and $Q$ in Eq.~(\ref{eq2}), and, of course, from Eq.~(\ref{eqdifus}), which describes the protein mass conservation. Before commenting on these differences, we explain the physical meaning of these four equations. The first one, Eq.~(\ref{eq1}), is the balance of stresses normal to the 
membrane, with $\nabla^2f_h$ and $-\sigma\nabla^2h$ the elastic stresses and $2\eta D_{zz}^\pm-p^{\pm}$ the stresses arising from the solvent. Equation~(\ref{eq2}) is the balance of stresses parallel to the membrane in each monolayer: the first two terms are the viscous stresses, the third term is the intermonolayer friction, the fourth term the gradient of two-dimensional pressure, the fifth term ($\propto\!Q$) is an extra intermonolayer stress and the last term is the solvent shear stress. Eq.~(\ref{eq3}) is the mass continuity equation and Eq.~(\ref{eqdifus}) is the protein conservation equation.

In order to grasp the differences with the dynamics of membranes deprived of spanning objects, we consider the limit where the proteins are located only in the upper monolayer, which corresponds to $s^+=1$ and $s^-\sim s\to\infty$. In this case $\delta b\to0$, $Qs^+\sim s^{-2}\to0$, $Qs^-\sim s^{-1}\to0$, and $S s^+\to 0$, and all the extra terms mentioned above vanish. These terms arise thus specifically from the spanning character of the proteins. As explained in the introduction, the increase of the intermonolayer friction coefficient $b$ arises from the dissipation $\propto\!1/\Gamma$ caused by the motion of the lipids relative to the proteins. Note that a similar conclusion was found in a simulation study~\cite{Khoshnood:2010aa}. More precisely, in the process of obtaining Eqs.~(\ref{eq1})--(\ref{eqdifus}), the protein currents $J_i^\pm$ are found to bear a term proportional to the relative barycentric velocity $v_i^+-v_i^-$. This comes in part from Eq.~(\ref{Jconstraint}). Thus the dissipation associated with diffusion, $J_i^{\pm}J_i^{\pm}/(2\Gamma)$, contains a term proportional to $(v_i^+-v_i^-)^2$ which increases $b$.
Note that in Eq.~(\ref{B}) the dependence on $c_0$ is hidden, since $\Gamma$ depends on $c_0$. Hence in the dilute limit $\delta b\sim c_0$ since $\Gamma\sim c_0$.  As for the stresses $\propto\!Q$, which are proportional to the gradient of the energy density $f_\phi^\pm$, they are elastic stresses transmitted by the spanning character of the proteins.

\section{Relaxation modes}
\label{Relax_modes}

The equations being linear, we may decompose all quantities into independent in-plane Fourier modes: $h(\bm x_{\perp},t)=\int d^2q(2\pi)^{-2} h_{\bm q}(\bm q,t)\exp(i\bm q\cdot\bm x_{\perp})$, etc. Let us define the average density $\bar\rho=\rho^+ + \rho^-$ and the density--difference
\begin{align}
\rho=\rho^+ - \rho^-.
\end{align}
Taking the Fourier transform of the membrane equations~(\ref{eq1}--\ref{eq3}) and eliminating the bulk and surface velocities yields a system of four equations that give $\dot h_{\bm q}$, $\dot\rho_{\bm q}$, $\dot{\bar\rho}_{\bm q}$ and $\dot\phi_{\bm q}$ as linear combinations of $h_{\bm q}$, $\rho_{\bm q}$, $\bar\rho_{\bm q}$ and $\phi_{\bm q}$. This procedure, involving heavy calculations, is detailed in the Appendix.

\subsection{Reduction to three dynamical variables}

We shall assume throughout typical parameter values: $\kappa^\pm\simeq10^{-19}$~J~\cite{Rawicz00BiophysJ}, $k^\pm\simeq0.1$~J/m$^2$~\cite{Rawicz00BiophysJ}, $e^\pm\simeq1$~nm, $\eta\simeq 10^{-3}$~J$\cdot$s/m$^3$, $\eta_s=10^{-9}$~J$\cdot$s/m$^2$~\cite{Honerkamp-Smith:2013aa}, $b\simeq 10^9$~J$\cdot$s/m$^4$~\cite{Merkel89JPhys,Pott02Europhys,Rodriguez-Garcia:2009aa}, and $\Dbare\simeq10^{-12}$~m$^2$/s for the diffusion coefficient of transmembrane proteins~\cite{Ramadurai09JACS}.  The coefficients $\alpha^\pm$, $\beta^\pm$, and $\lambda^\pm$ are expected to be of order unity~\cite{notealpha}.

\begin{figure}
\includegraphics[width=.92\columnwidth]{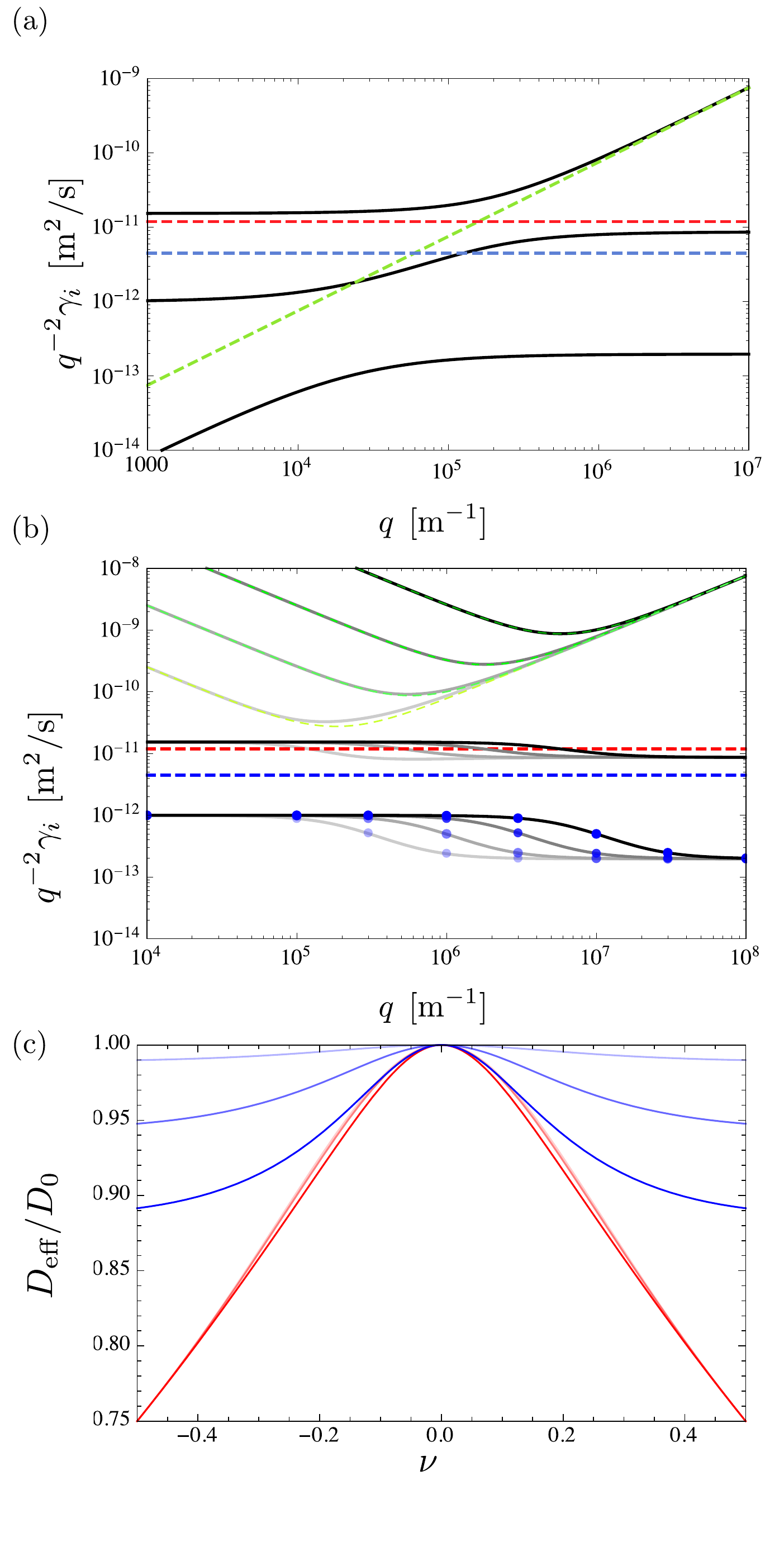}
\caption{\label{figs:Fig2}(Color online.) Relaxation rates and effective diffusion coefficient for asymmetric proteins.  All rates are normalized by $q^2$. (a) Zero tension case. The black lines show the three relaxation rates $\gamma_1>\gamma_2>\gamma_3$ versus wavevector $q$. The dashed lines show the $h$-like rate $\MA_{11}$ in green, the $\rho$-like rate $\MA_{22}$ in red and the $\phi$-like rate $\MA_{33}$ in blue. Besides the parameters given in the text, $b=5\times 10^8$~J$\cdot$s/m$^4$, $\nu=0.25$, $\hat e=0.1$~nm, $\hat\alpha=1$, $\hat\beta=1$, $\hat\lambda=1$. (b) Same plots in the finite tension case, $\sigma\ge\sigma_{\textrm{c}}$. Four values of $\sigma$ are shown: $\sigma=10^{-5}$~N/m (black), $\sigma=10^{-6}$~N/m (dark gray or dark green), $\sigma=10^{-7}$~N/m (gray or green), and $\sigma=10^{-8}$~N/m (light gray or light green). The dots indicate the effective diffusion coefficient $D_\mathrm{eff}$; it is indistinguishable from the slowest normalized rate.
(c) Asymptotic effective diffusion coefficients $\Deff^0$ for $q\to0$ (three upper blue curves) and $\Deff^\infty$ for $q\to\infty$ (three lower red curves), versus asymmetry parameter $\nu$.  In each group, from top to botton: $b=10^9$~J$\cdot$s/m$^4$, $b=5\times 10^9$~J$\cdot$s/m$^4$ and $b=10^{10}$~J$\cdot$s/m$^4$. The other specific parameters are $\hat e=\hat\alpha=\hat\beta=0$ and $\sigma=10^{-6}$~N/m. (For interpretation of the references to color, the reader is referred to the web version of this paper.)
}
\end{figure}

As in the case of protein-free membranes~\cite{Seifert93EPL}, it turns out, given the parameter values given above, that the variable $\bar\rho_{\bm q}$ is always much more rapid than $h_{\bm q}$, $\rho_{\bm q}$ and $\phi_{\bm q}$. It can therefore be eliminated by setting the right-hand-side of Eq.~(\ref{eqro}) equal to zero and solving for $\bar\rho_{\bm q}$. One then obtains a reduced dynamical system of the form 
\begin{align}
\begin{pmatrix}
\dot h_{\bm q}\\
\dot\rho_{\bm q}\\
\dot\phi_{\bm q}
\end{pmatrix}
=\mathsf{A}
\begin{pmatrix}
h_{\bm q}\\
\rho_{\bm q}\\
\phi_{\bm q}
\end{pmatrix}.
\label{eq:hrhophiDynamics}
\end{align}
In the general case, the coefficients of the matrix $\mathsf{A}$ are too cumbersome to be given explicitly.  Figures~\ref{figs:Fig2}a and \ref{figs:Fig2}b show examples of the three relaxation rates $\gamma_1>\gamma_2>\gamma_3$ given by the negative of the eigenvalues of $\mathsf{A}$, for different membrane tensions and degree of protein asymmetry. Note that they never cross.

To simplify the expression for the elements of $\mathsf{A}$, we consider the case of a weak asymmetry which we assume to be present only in the curvature coupling constants. Without loss of generality, it can be expressed in terms of a small parameter $\nu$, by setting
$1/s^\pm=1/2\pm\nu$, $e^\pm=e\pm\nu\hat e$, $\beta^\pm=\beta\pm\nu\hat\beta$, $\lambda^\pm=\lambda\pm\nu\hat\lambda$, and $\kappa^\pm=\kappa$, $k^\pm=k$, $r_0^\pm=r_0$, $\alpha^\pm=\alpha$. We then obtain
\begin{align}
\MA=\MA^{(0)}+\nu\MA^{(1)}+O(\nu^2)
\end{align}
with 
\begin{align}
\MA^{(0)}=
\begin{pmatrix}
\displaystyle-\frac{\tilde\kappa q^3+\sigma q}{4\eta}
& \displaystyle\frac{ke(1 -c_0')q}{4\eta}
& 0\medskip\\
\displaystyle\frac{ke(1 -c_0')q^4}{B+\eta q+\eta_s q^2} 
& \displaystyle-\frac{k(1-c_0'')q^2}
{2(B+\eta q+\eta_s q^2)} 
& 0\medskip\\
0 & 0 
& \displaystyle-\frac{\Gamma k\alpha q^2}{r_0^2}
\end{pmatrix}.
\end{align}
where $B=b+\delta b$, $\tilde\kappa=\kappa+2ke^2$, $\eta_s=\eta_2+\frac{1}{2}\lambda_2$, $c_0'=c_0(\beta+\lambda)$, $c_0''=2\beta c_0-c_0^2(\alpha+\beta^2)$, and
\begin{align}
&\MA^{(1)}_{13}=\frac{k e_1}{2\eta}\,q\,,\quad
\MA^{(1)}_{23}=\frac{2\tilde\beta k}{B+\eta q+\eta_s q^2}\,q^2,
\\
&\MA^{(1)}_{31}=\left(\frac{\Gamma e_1}{r_0^2}+\frac{2e c_0 (1-c_0')}{B+\eta q+\eta_s q^2}\right)kq^4,
\\
&\MA^{(1)}_{32}=\left(\frac{(2c_0\alpha+\tilde\beta)\Gamma}{r_0^2}-\frac{c_0(1-c_0'')}{B+\eta q+\eta_s q^2}\right)kq^2,
\end{align}
with $e_1=(2\beta+\hat\beta+2\lambda+\hat\lambda)e+\lambda \hat e$ and $\tilde\beta=(\beta+\frac{1}{2}\hat\beta)(c_0\beta-1)$, the other elements of $\MA^{(1)}$ being zero. 

It can be first noticed that, for symmetrical proteins ($\nu=0$), the matrix $\MA=\MA^{(0)}$ is block-diagonal and $h$ and $\rho$ are coupled exactly as in protein-free membranes, but with renormalized coefficients~\cite{compareSeifert}. 

Next we compare the basic relaxation rates in this problem, given by the diagonal elements of the matrix~$\MA$. Although they do not coincide in general with the actual rates $\gamma_i$, they provide useful informations. First, the $h$-like relaxation rate, associated with $\MA^{(0)}_{11}$, is $(\tilde{\kappa}q^3+\sigma q)/(4\eta)$ (green dashed line in Fig.~\ref{figs:Fig2}a,b). Second, the $\rho$-like rate, associated with $\MA^{(0)}_{22}$ and originating from the relative monolayer movement, is $\simeq \!k q^2/B$ (red dashed line in Fig.~\ref{figs:Fig2}a,b). Here, we have assumed $c''_0=\mathcal{O}(1)$ and $B\gg\eta q+\eta_s q^2$, which we shall assume throughout since it holds typically for $q^{-1}\gtrsim1$~nm, given the orders of magnitude listed above. The ratio of the $h$-like rate to the $\rho$-like rate is minimum where the green dashed line of Fig.~\ref{figs:Fig2}a ($\sigma=0$) or \ref{figs:Fig2}b ($\sigma \neq 0$) attains its minimum, \textit{i.e.}, at
\begin{align}
q_{\textrm{c}}=\sqrt{\sigma/\tilde\kappa}.
\end{align}
Comparing the $h$-like rate and the $\rho$-like rate yields the crossover tension
\begin{align}
\sigma_{\textrm{c}}=\frac{4}{\tilde{\kappa}}\left(\frac{k\eta}{B}\right)^2\lesssim 10^{-8}~\mathrm{N/m}
\end{align}
that separates the vanishing tension regime (Fig.~\ref{figs:Fig2}a) from the finite tension regime (Fig.~\ref{figs:Fig2}b); it is defined by the condition that for $\sigma>\sigma_{\textrm{c}}$ the $h$-like  rate is always larger that the $\rho$-like rate (as in Fig.~\ref{figs:Fig2}b).

Finally, the $\phi$-like rate, associated with $\MA^{(0)}_{33}$, is $\Dbare q^2$ (blue dashed line in Fig.~\ref{figs:Fig2}a,b), where 
\begin{equation}
\Dbare=\frac{k\Gamma\alpha}{r_0^2}\,.
\label{Dbare}
\end{equation}
The ratio of the $\phi$-like rate to the $\rho$-like rate (that both have a diffusive character $\propto\!q^2$) is $D_0B/k$; this leads us to introduce the small dimensionless parameter:
\begin{align}
\zeta=\frac{\Dbare b}{k}\approx10^{-2}.
\end{align}
We thus expect in general $\phi_{\bm q}$ to be the slowest variable. 

With the above considerations, we can write the renormalized intermonolayer friction coefficient [Eq. (\ref{B})], for symmetric membranes, as
\begin{equation}
B=b\left(1+\frac{\alpha c_0^2}{2\zeta}\right)\,.
\end{equation}
Thus, the slower the diffusion of the proteins, the larger the renormalization of the intermonolayer friction coefficient. Since we expect $\alpha\approx1$ in the absence of specific interactions among the proteins~\cite{notealpha}, and assuming typically $c_0\approx0.3$, the renormalization is large, owing to the smallness of $\zeta$.  
Thus, transmembrane proteins are expected to have a significant influence on the intermonolayer movement, and hence, on the relaxation of membrane bending fluctuations~\cite{Seifert93EPL,Pott02Europhys,Shkulipa:2006aa,Rodriguez-Garcia:2009aa}.

The actual relaxation rates $\gamma_i$ can be interpreted as follows. The quickest rate, $\gamma_1$, always coincides (more or less perfectly) with one of the three basic relaxation rates (Fig.~\ref{figs:Fig2}a,b). For instance, at high $q$, the quickest rate (upper black line) coincides with the $h$-like rate (green dashed line). This means that $h_{\bm q}$ relaxes very quickly with $\rho_{\bm q}$ and $\phi_{\bm q}$ ``frozen''. The medium rate (intermediate black line), $\gamma_2$, is mainly parallel but below another one of the three basic rates, indicating the second--quickest variable. For instance, at high $q$, this is the $\rho$-like rate, which implies that after $h_{\bm q}$ has relaxed, $\rho_{\bm q}$ is relaxing with both $h_{\bm q}$ ``slaved'' and $\phi_{\bm q}$ still ``frozen''. Then, the slowest rate, $\gamma_3$, is parallel and below the slowest basic rate, indicating the slowest variable. For instance, at high $q$, $\phi_{\bm q}$ is the slowest variable and it relaxes with $h_{\bm q}$ and $\rho_{\bm q}$ ``slaved''. This analysis applies wherever the three rates are well separated, outside the crossover regions. For instance, at zero tension and low $q$ (Fig.~\ref{figs:Fig2}a) the sequence from quickest to slowest relaxing variables is $\rho_{\bm q}$, $\phi_{\bm q}$ and $h_{\bm q}$. Actually, $h_{\bm q}$ is the slowest variable only for $\sigma=0$ and low $q$. At finite tensions, $\sigma>\sigma_{\textrm{c}}$, the slowest variable is always $\phi_{\bm q}$ (Fig.~\ref{figs:Fig2}b) and its relaxation rate steps downwards for $q>q_{\textrm{c}}$; the slowing down of protein diffusion at large $q$ has long been recognized as resulting from the coupling to membrane curvature~\cite{Leibler:1986}.

\subsection{Effective protein dynamics at finite\newline membrane tension}

From Fig.~\ref{figs:Fig2}b and the discussion just above, we see that for ordinary membranes, for which $\sigma\gtrsim 10^{-8}$~N/m~$\gg\sigma_{\textrm{c}}$, the relaxation rate $\gamma_1$ coincides with the relaxation rate of $h_{\bm q}$, and is at least an order of magnitude greater than the other two. Therefore, $h_{\bm q}$ adiabatically follows the dynamics of $\rho_{\bm q}$ and $\phi_{\bm q}$, and by setting $\dot{h}_{\bm q}=0$ in Eq.~\eqref{eq:hrhophiDynamics} we are able to reduce the dynamical problem to two variables, governed by the matrix
\begin{align}
\MB&=
\begin{pmatrix}
\MA_{22}-\displaystyle{\frac{\MA_{21}\MA_{12}}{\MA_{11}} }& \MA_{23}-\displaystyle{\frac{\MA_{21}\MA_{13}}{\MA_{11}} } \medskip \\
\MA_{32}-\displaystyle{\frac{\MA_{31}\MA_{12}}{\MA_{11}} }& \MA_{33}-\displaystyle{\frac{\MA_{31}\MA_{13}}{\MA_{11}} }
\end{pmatrix}\,.
\end{align}
The two relaxation rates (negative of eigenvalues) $\Gamma_1>\Gamma_2$ of $\MB$ are not always well separated, and one cannot eliminate adiabatically $\rho_{\bm q}$ in order to get the dynamics of $\phi_{\bm q}$. Instead, keeping only the slower rate, one has $\phi_{\bm q}(t)\sim\exp[-q^2\Deff(q) t]$, where $\Deff=\Gamma_2/q^2$ and $\Gamma_2=-\frac{1}{2}\{\MB_{11}+\MB_{22}+[(\MB_{11}-\MB_{22})^2+4 \MB_{12}\MB_{21}]^{1/2}\}$. The precise form of $\Deff(q)$, the effective diffusion coefficient of the proteins, is quite complicated for arbitrary $q$, but can be calculated numerically, as shown in Fig.~\ref{figs:Fig2}c. Its coincidence with $q^{-2}\gamma_3$ justifies the adiabatic approximation for $h_q$.

We consider now two limiting forms of $\Deff(q)$. First, for $q\to 0$, the asymptotic effective diffusion coefficient, $\Deff^0$, is given at first order in the small parameter $\zeta$ and to $\mathcal{O}(\nu^2)$ by
\begin{equation}
\frac{ \Deff^0}{\Dbare}\simeq
1-2\nu^2\zeta\,
\frac{(2\beta +\hat\beta)^2}
{\alpha(\beta c_0-1)^2}\,.
\end{equation}   
Since $\zeta\propto b$, we find that the protein diffusion coefficient at long wavelengths is, for asymmetric  bilayers, reduced by the intermonolayer friction $b$ (see Fig.~\ref{figs:Fig2}c): this effect arises from the coupling between $\phi$ and $\rho$ in the free energy terms $\beta^{\pm}\phi^{\pm}{}^2$ through Eq.~\eqref{eq:MonoProtDensITOBilayerDens}.  
   
Second, for $q\gg q_{\rm{c}}$, we write
\begin{equation}
\Deff^{\infty}=\Deff^0+\Delta D\,,
\end{equation}
where in general $\Delta D$ is very complicated.  However, $\Delta D$ vanishes for symmetric membranes ($\nu=0$) and no protein--membrane curvature coupling ($\lambda=\hat{\lambda}=0$), in agreement with Ref.~\cite{Leibler:1986}.  For the simple asymmetric case $\hat{e}=\hat{\lambda}=0$ and $c_0\to 0$, expanding again to $\mathcal{O}(\nu^2)$ and to first order in $\zeta$ yields
\begin{equation}
\frac{\Delta D}{\Dbare}=-8\nu^2\,\frac{ke^2}{\kappa}\frac{\lambda^2}{\alpha}\left[1+\frac{2\zeta}{\lambda}
   \left(2 \beta 
   +\hat{\beta }  -2 \frac{ k e^2}{\kappa}
  \right)  \right]\,.
\end{equation}
Thus, the effect of intermonolayer friction, $b\propto\zeta$, is a secondary effect compared to the dominant role played by the protein-membrane curvature coupling $\lambda$ (see Fig.~\ref{figs:Fig2}c). 

\section{Diffusion of a concentrated protein spot}
\label{Diffusion_spot}

As an application of the above formalism, we consider the diffusion of a concentrated spot of asymmetric transmembrane proteins. The relaxation of the protein density field, $c(\bm {x}_{\perp},t)$, is coupled to the membrane height, $h(\bm{x}_{\perp},t)$, and to the density--difference $\rho(\bm{x}_{\perp},t)$.
Defining the state of the protein--membrane system by the column matrix
$\mathsf{X}(\bm{x}_{\perp}, t)= (h(\bm{x}_{\perp}, t),\rho(\bm{x}_{\perp}, t),\phi(\bm{x}_{\perp}, t))^{T}$, we ask how does $\mathsf{X}(\bm{x}_{\perp}, t)$ evolve given that proteins are initially concentrated in a circular region of radius $w$ in a flat, equilibrium membrane:
\begin{equation}
\mathsf{X}(\bm{x}_{\perp}, 0)=\phi_0\,e^{- r^2/(2\,w^2)}\,(0, 0, 1)^T\,,
\end{equation}
To find out, we work in Fourier space. The time evolution of the Fourier 
transform is determined by solving $\dot{\mathsf{X}}(\bm{q},t) = \mathsf{A}\,\mathsf{X}(\bm{q},t)$ for each $\bm{q}$, subject to the 
initial condition:
\begin{align}
\mathsf{X}(\bm q,0)&=\int\!d^2r\,\mathsf{X}(\bm{x}_{\perp},0)\,e^{-i\bm{q}\cdot\bm{x}_{\perp}} \nonumber\\&
= 2\pi w^2\,\phi_0\,
e^{-\frac{1}{2}w^2q^2}
\,(0, 0, 1)^T\,.
\end{align}
We denote $\mathsf{S}(q)$ the matrix whose columns are the three eigenvectors of $\mathsf{A}$; therefore $\mathsf{A}_d=\mathsf{S}^{-1}\mathsf{A}\mathsf{S}$ is a diagonal matrix whose elements are the negative of the relaxation rates $\gamma_i(q)$. With a change of basis, we define $\mathsf{X}'(\bm{q},t)=\mathsf{S}^{-1}\mathsf{X}(\bm{q},t)$, which satisfies 
$\dot{\mathsf{X}'}(\bm{q},t) = \mathsf{A}_d \mathsf{X}'(\bm{q},t)$, with solution $\mathsf{X}'(\bm{q},t)=e^{\mathsf{A}_d t}\,\mathsf{X}'(\bm{q},0)$. Therefore, we obtain $\mathsf{X}(\bm{q},t)=\mathsf{S}\,e^{\mathsf{A}_d t}\mathsf{S}^{-1}\,\mathsf{X}(\bm q,0)$.
As a result, the solution in real space is reconstructed, yielding
\begin{align}
\mathsf{X}(\bm x_\perp,t)&=\int \frac{d^2 q}{(2\pi)^2} \,\mathsf{S}\,e^{\mathsf{A}_d t}\mathsf{S}^{-1}\,\mathsf{X}(\bm q,0)\,e^{i\bm{q}\cdot\bm{x}_{\perp}}
\nonumber \\
&= \int_0^{\infty} \frac{dq}{2\pi} \,q J_0(q\,r)\,\mathsf{S}\,e^{\mathsf{A}_d t}\mathsf{S}^{-1}\,\mathsf{X}(\bm q,0)\,,
\end{align}
and thus,
\begin{align}
\phi(r,t)=&
\phi_0 w^2\sum_i\int_0^{\infty}\!dq\,q J_0(qr)
\times\nonumber\\&\times
\mathsf{S}_{3i}(q)\,
e^{-\frac{1}{2}w^2 q^2-\gamma_i(q) t}\,
\mathsf{S}^{-1}_{i3}(q).
\end{align}
We have used the symmetry of revolution of the problem to carry out the angular integration, which yields the Bessel function of zeroth order $J_0(qr)$, where $r=|\bm x_\perp|$. We therefore see that the diffusion of the protein spot, viewed in real space, is governed not only by the bare diffusion constant $\Dbare$, but more generally by all the three rates~$\gamma_i$.

\begin{figure}[h]
\includegraphics[width=.95\columnwidth]{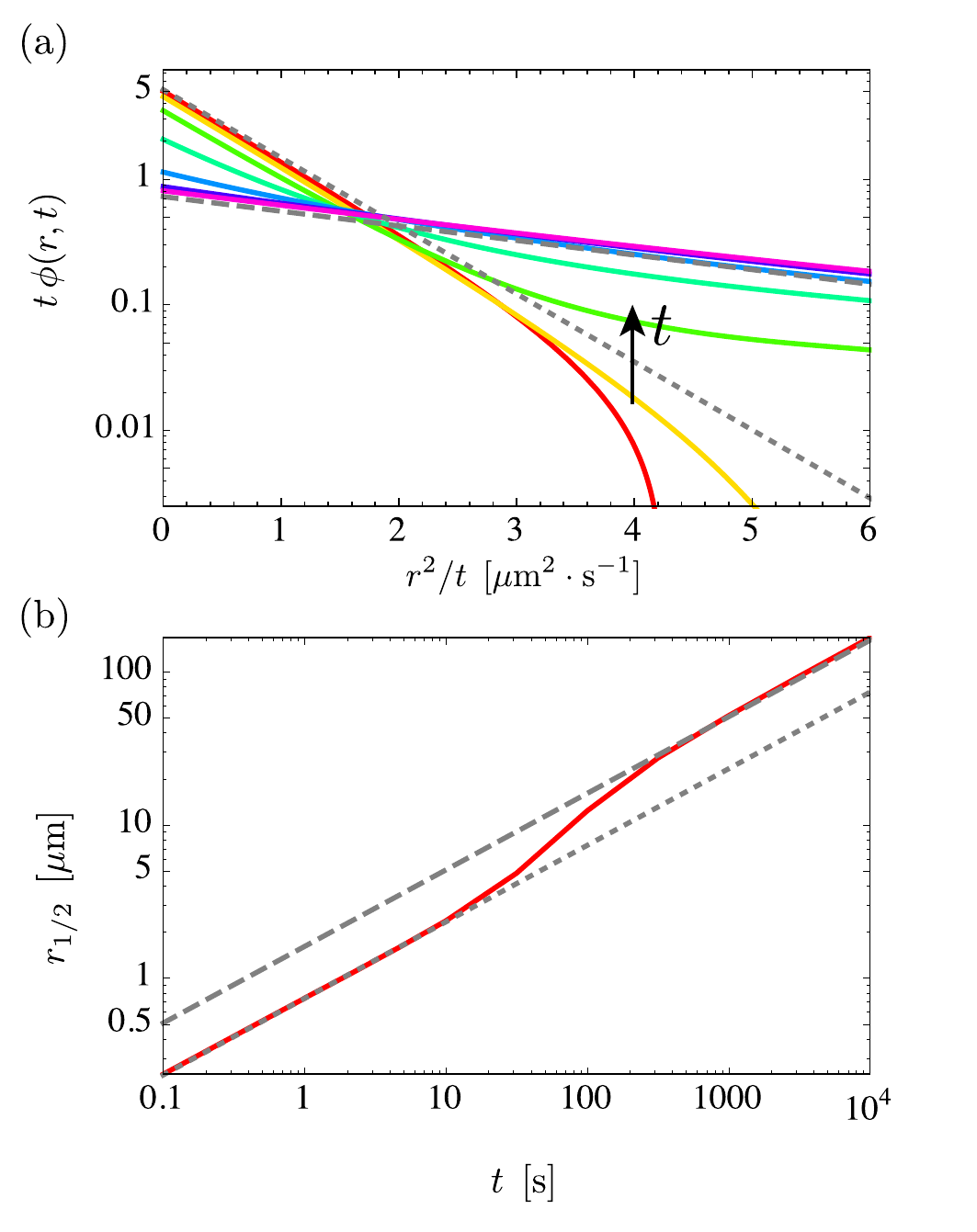}
\caption{\label{figs:Fig3}(Color online.)  
Diffusion of a concentrated protein spot in a tensed membrane. (a) Logarithmic plot of $t\,\phi(r,t)$ normalized by $2\Dbare/(\phi_0 w^2)$ versus $r^2/t$ for times ranging from $t=1$ s to $t=1000$ s, with two profiles per time decade (red to fuchsia, from bottom to top). Short and long time behaviors are given by dotted and dashed lines, controlled by $\Deff^\infty$ and $\Deff^0$ (see text). (b) Half width at half maximum, $r_{1/2}$, as a function of $t$.  
Short and long time behaviors are given by dotted and dashed lines, controlled by $\Deff^\infty$ and $\Deff^0$  (see text). In addition to the parameter values given in text, $\sigma=10^{-8}$~N/m, $b=5\times 10^{9}$~Js/m$^4$, $\hat e=0.1$~nm, $\hat\alpha=1$, $\hat\beta=1$, $\hat\lambda=1$, $\nu=0.25$, and $w=10^{-8}$~m.
}
\end{figure}

The time evolution of the protein concentration spot is described in Fig.~\ref{figs:Fig3}. The $q$-dependence of the effective diffusion coefficient $\Deff(q)$, as seen in Fig.~\ref{figs:Fig2}b, is reflected in the evolution of the real space protein mass fraction $\phi(r,t)$.  As shown in Fig.~\ref{figs:Fig3}a, at short times (but still $\Dbare t> w^2$) and for $r^2\lesssim 2\Dbare t$, the protein diffusion is controlled by $\Deff^{\infty}$.
Indeed, for times such that 
$q_{\textrm{c}} \sqrt{\Dbare t}\ll 1$, 
\begin{equation}
\phi(r,t)\simeq  \frac{\phi_0 w^2\,a_3^{\infty}}{2\Deff^\infty t}e^{-\frac{r^2}{4\Deff^\infty t}}\,,\vspace{0.25cm}
\end{equation}
where $a_3^{\infty}\equiv \lim_{q\to \infty} \mathsf{S}_{33}\mathsf{S}^{-1}_{33}$.  Note that, for large $r$, $\phi$ oscillates slightly about zero, due to the coupling with membrane curvature for large $q$; in Fig.~\ref{figs:Fig3}a this is seen as a divergence in the logarithm of $t\,\phi(r,t)$.  

In contrast, at long times ($q_{\textrm{c}} \sqrt{\Dbare t}\gg 1$), the protein 
diffusion is controlled by $\Deff^{0}$, and 
\begin{equation}
\phi(r,t)\simeq  \frac{\phi_0 w^2\,a_3^{0}}{2\Deff^0 t}e^{-\frac{r^2}{4\Deff^0 t}}\,,
\end{equation}
where $a_3^{0}\equiv \lim_{q\to 0} \mathsf{S}_{33}\mathsf{S}^{-1}_{33}$.
These behaviors are confirmed by Fig.~\ref{figs:Fig3}b, showing the half width at half maximum of $\phi$, denoted $r_{1/2}$, versus $t$.  
 For short times, $r_{1/2}$ tends to $2\sqrt{\Deff^{\infty}\,t\ln 2}$, while for long times, $r_{1/2}$ tends to $2\sqrt{\Deff^{0}\,t\ln 2}$.
Thus, the coupling of protein diffusion to membrane curvature is felt at short times (through $\Deff^{\infty}$), whereas the coupling to in-plane motion is felt at long times (through $\Deff^0$).

\section{Conclusion and Perspectives}
\label{Conclusions}

To summarize, we developed in this paper a generic formalism, based on Onsager's variational principle, that allowed us to describe the dynamics of lipid bilayers containing transmembrane proteins of arbitrary shape. This approach, valid on length scales larger than the mean inter-protein distance, reveals that the presence of proteins that span the bilayer opposes monolayer sliding, and hence significantly increases the intermonolayer friction coefficient. The correction to this coefficient is shown to vary in inverse proportion to the protein's mobility. Experimentally, transmembrane protein-enhanced intermonolayer friction has recently been invoked to explain observations of slow force relaxation on membrane tubes extracted from Giant Unilamellar Vesicles~\cite{Campillo:2013aa}.  Applying our formalism to the simple situation where protein enrichment is localized initially in a circular region within a flat membrane, we show that for practically all accessible membrane tensions, protein density is the slowest relaxing variable, and the spreading of the protein spot follows an anomalous diffusion behavior. Precisely, at short and long times, the spreading passes from a normal diffusion regime at short times to another normal diffusion regime (\textit{i.e.} with another diffusive coefficient) at long times. To our knowledge, diffusion of transmembrane proteins has been probed only at very low concentrations~\cite{Quemeneur14}. The anomalous diffusion behavior expected at larger protein concentrations could be tested experimentally: the initial patch of proteins could be realized either by applying a magnetic field on proteins with grafted magnetic particles, or by coalescing a small vesicle with high concentration of proteins and a large vesicle with moderate concentration of proteins \cite{Bassereau-Mangenot}. Some proteins, like Gramicidin, can exist in the form of monomers located separately in the two monolayers, or in the form of a dimer spanning the whole bilayer~\cite{GoulianBJ98}. In light of our model, it would be interesting to test experimentally the effect of this dimerization on the effective intermonolayer friction coefficient.
\section*{Acknowledgments}
We thank S. Mangenot and P. Bassereau for useful discussions.  J.-B. Fournier acknowledges financial support from the French A\-gen\-ce Nationale de
la Recherche (Contract No.\ ANR-12-BS04-0023-MEMINT).
%

\section*{Appendix}
\renewcommand{\thesubsection}{\Alph{subsection}}
\subsection{Dynamical equations for the membrane in Fourier space}

In reciprocal space, Eqs.~(\ref{eq1}--\ref{eqdifus}) take the form:
\begin{widetext}
\begin{align}
&-q^2\!\sum_{\epsilon=\pm}\left(
f_{h\rho}^\epsilon\,\rho_{\bm q}^\epsilon
+f_{h\phi}^\epsilon\,\phi_{\bm q}^\epsilon
+f_{hh}^\epsilon\, h_{\bm q}
\right)
+\sigma q^2h_{\bm q} +4\eta q \dot h_{\bm q}=0\,,
\label{A1}
\\&
2\eta_s q^2 v^\pm_{\bm q}\pm B(v^+_{\bm q} - v^-_{\bm q})
+iq\left(
f_{\rho\rho}^\pm\,\rho^\pm_{\bm q}
+f_{\rho\phi}^\pm\,\phi^\pm_{\bm q}
+f_{\rho h}^\pm\, h_{\bm q}
\right)
\nonumber\\&~~~~~~~~~~~~~~~~~~~~~~
\mp i q Q\!\sum_{\epsilon=\pm} \epsilon s^\epsilon r_0^{-\epsilon}
\left(
f_{\phi\rho}^\epsilon\,\rho^\epsilon_{\bm q}
+f_{\phi\phi}^\epsilon\,\phi^\epsilon_{\bm q}
+f_{\phi h}^\epsilon\, h_{\bm q}
\right)+2\eta q\, v^\pm_{\bm q} = 0\,,
\label{A2}
\\&
\dot \rho_{\bm q}^\pm+i q v^\pm_{\bm q}=0\,,
\\&
\dot \phi+S q^2\!\sum_{\epsilon=\pm} s^{-\epsilon} r_0^{-\epsilon}
\left(
f_{\phi\rho}^\epsilon\,\rho^\epsilon_{\bm q}
+f_{\phi\phi}^\epsilon\,\phi^\epsilon_{\bm q}
+f_{\phi h}^\epsilon\, h_{\bm q}
\right)
-iqT(v^+_{\bm q}-v^-_{\bm q})=0\,,
\end{align}
where $v^\pm_{\bm q}$ is the component of the velocity parallel to $\bm q$.
The above coefficients $f^\pm_{ij}$, where $(i,j)\in\{\rho,\phi,h\}^2$, are defined through the reciprocal space relations: $f^\pm_\rho=
f^\pm_{\rho\rho}\,\rho^\pm_{\bm q}+f^\pm_{\rho\phi}\,\phi_{\bm q}^\pm+f^\pm_{\rho h}\,h_{\bm q}$, $f^\pm_\phi=f^\pm_{\phi\rho}\,\rho^\pm_{\bm q}+f^\pm_{\phi\phi}\,\phi^\pm_{\bm q}+f^\pm_{\phi h}\,h_{\bm q}$, and $f^\pm_h=
f^\pm_{h\rho}\,\rho^\pm_{\bm q}+f^\pm_{h\phi}\,\phi^\pm_{\bm q}+f^\pm_{hh}\,h_{\bm q}$. Explicitly, they read
\begin{align}
\begin{tabular}{lll}
$f^\pm_{\rho\rho}=k^\pm$,
&
$f^\pm_{\rho\phi}=k^\pm\beta^\pm$,
&
$f^\pm_{\rho h}=\mp e^\pm k^\pm q^2,$
\\
$f^\pm_{\phi\rho}=k^\pm\beta^\pm$,
&
$f^\pm_{\phi\phi}=k^\pm\alpha^\pm+k^\pm{\beta^\pm}^2$,
&
$f^\pm_{\phi h}=\mp e^\pm k^\pm(\lambda^\pm+\beta^\pm)q^2,$
\\
$f^\pm_{h\rho}=\pm e^\pm k^\pm$,
&
$f^\pm_{h\phi}=\pm e^\pm k^\pm(\lambda^\pm+\beta^\pm)$,
&
$f^\pm_{hh}=-\frac{1}{2}(\kappa^\pm+2k^\pm{e^\pm}^2)q^2.$
\end{tabular}
\end{align}
The term $4\eta q\dot h_{\bm q}$ in Eq.~(\ref{A1})
is the Fourier transform of the bulk stress $p^+-p^--2\eta(D_{zz}^+-D_{zz}^-)$ evaluated at $z=0$; the term $2\eta_s q^2 v^\pm_{\bm q}$ in Eq.~(\ref{A2}), where $\eta_s=\eta_2+\frac{1}{2}\lambda_2$, is the Fourier transform of $-2\eta_2\partial_j d_{ij}^\pm -\lambda_2 \partial_i d_{jj}^\pm$; finally, the term $2\eta qv^\pm_{\bm q}$ in Eq.~(\ref{A2}) is the Fourier transform of the bulk stress $\mp2\eta D^\pm_{zi}$ evaluated in $z=0$. For a detailed derivation of those contributions, see, e.g., Sec.~8 of Ref.~\cite{Fournier15IJNM}.

Expressing $\phi^\pm_{\bm q}$ in terms $\phi_{\bm q}$, $\rho_{\bm q}$ and $\bar\rho_{\bm q}$ using Eq.~(\ref{eq:MonoProtDensITOBilayerDens}) and eliminating the velocities $v^\pm_{\bm q}$ yields the system of equations given in the body of the text: 
\begin{align}
\label{mat44debut}
4\eta\dot h&=(f^+_{hh}+f^-_{hh}-\sigma)qh
+\frac{1}{2}(f_{h\rho}^+ -f_{h\rho}^- +2Y^-f_{h\phi}^- -2Y^+f_{h\phi}^+)q\rho
+\frac{1}{2}(f_{h\rho}^+ + f_{h\rho}^-)q\bar\rho
\nonumber\\&
+(X^+ f_{h\phi}^+ + X^- f_{h\phi}^-)q\phi,
\\
2(B+\eta q+\eta_s q^2)\dot\rho&=(f_{\rho h}^- - f_{\rho h}^+ -2Qs^-r_0^+ f_{\phi h}^- + 2Qs^+r_0^- f_{\phi h}^+)q^2h
\nonumber\\&
-\frac{1}{2}[
f_{\rho\rho}^- + f_{\rho\rho}^+
-2Y^-f_{\rho\phi}^- -2Y^+f_{\rho\phi}^+
-2Qr_0^+s^-(f_{\phi\rho}^- - 2Y^-f_{\phi\phi}^-)
 -2Qr_0^-s^+(f_{\phi\rho}^+ - 2Y^+f_{\phi\phi}^+)] q^2\rho
\nonumber\\&
+\frac{1}{2}(
f_{\rho\rho}^- - f_{\rho\rho}^+ -2Qr_0^+s^-f_{\phi\rho}^-
+2Qr_0^-s^+f_{\phi\rho}^+
)q^2\bar\rho
\nonumber\\&
+q^2(
X^-f_{\rho\phi}^- - X^+f_{\rho\phi}^+
-2X^-Qr_0^+s^-f_{\phi\phi}^-
+2X^+Qr_0^-s^+f_{\phi\phi}^+
)\phi,
\\
\label{eqro}
2(\eta+\eta_sq)\dot{\bar\rho}&=
-(f_{\rho h}^+ + f_{\rho h}^-)qh
+\frac{1}{2}
(
f_{\rho\rho}^- - f_{\rho\rho}^+
-2Y^- f_{\rho\phi}^- +2Y^+ f_{\rho\phi}^+
)q\rho
-\frac{1}{2}(f_{\rho\rho}^+ + f_{\rho\rho}^-)q\bar\rho
\nonumber\\&
-(X^+f_{\rho\phi}^+ + X^-f_{\rho\phi}^-)q\phi,
\\
\dot\phi&=\frac{1}{2}\left(-2SM_{41}+\frac{TN_{41}}{B+\eta q+\eta_s q^2}\right)q^2h
+\frac{1}{4}\left(-2SM_{42}+\frac{TN_{42}}{B+\eta q+\eta_s q^2}\right)q^2\rho
\nonumber\\&
+\frac{1}{4}\left(-2SM_{43}+\frac{TN_{43}}{B+\eta q+\eta_s q^2}\right)q^2\bar\rho
+\frac{1}{2}\left(-2SM_{44}+\frac{TN_{44}}{B+\eta q+\eta_s q^2}\right)q^2\phi,
\label{mat44fin}
\end{align}
where $X^\pm=(r_0^++r_0^-)/(r_0^\pm s^\pm)$ and $Y^\pm=(r_0^\mp c_0)/(r_0^\pm s^\pm)$ are the coefficients appearing in Eq.~(\ref{eq:MonoProtDensITOBilayerDens}), and the coefficients $M_{4i}$ and $N_{4i}$ are given by
\begin{align}
M_{41}&=r_0^-s^-f_{\phi h}^+ + r_0^+s^+f_{\phi h}^-,\\
M_{42}&=2r_0^+s^+(f_{\phi\rho}^--2Y^-f_{\phi\phi}^-)
-2r_0^-s^-(f_{\phi\rho}^+-2Y^+f_{\phi\phi}^+),
\\
M_{43}&=r_0^+s^+f_{\phi\rho}^-
+r_0^-s^-f_{\phi\rho}^+,
\\
M_{44}&=
r_0^-s^-X^+f_{\phi\phi}^+
+ r_0^+s^+X^-f_{\phi\phi}^-,
\\
N_{41}&=f_{\rho h}^+ - f_{\rho h}^-
+2Q(r_0^+s^-f_{\phi h}^-
-r_0^-s^+f_{\phi h}^+),
\\
N_{42}&=f_{\rho\rho}^- + f_{\rho\rho}^+
-2Y^-(f_{\rho\phi}^--2Qr_0^+s^-f_{\phi\phi}^-)
-2Y^+(f_{\rho\phi}^+-2Qr_0^-s^+f_{\phi\phi}^+)
-2Q(r_0^+s^-f_{\phi\rho}^- + r_0^-s^+f_{\phi\rho}^+),
\\
N_{43}&=f_{\rho\rho}^+ - f_{\rho\rho}^-
+2Q(r_0^+s^-f_{\phi\rho}^- - r_0^-s^+f_{\phi\rho}^+),
\\
N_{44}&=X^+f_{\rho\phi}^+ - X^-f_{\rho\phi}^-
+2Q(
r_0^+s^-X^-f_{\phi\phi}^-
- r_0^-s^+X^+f_{\phi\phi}^+
).
\end{align}
\end{widetext}

\bibliography{Hydrobiprot3}

\begin{mcitethebibliography}{54}
\bibitem[jus()]{justif1}
The term $\sim\!\nabla^2h$ is discarded as a boundary term. The terms
  $\sim\!\rho^\pm$ and $\sim\!\phi^\pm$ are discarded as both $\int
  d^2x_\perp\, \rho^\pm$ and $\int d^2x_\perp\, (1+\rho^\pm)\phi^\pm$ are
  constant due to the mass conservation of lipids and proteins.\relax
\mciteBstWouldAddEndPunctfalse
\mciteSetBstMidEndSepPunct{\mcitedefaultmidpunct}
{}{\mcitedefaultseppunct}\relax
\EndOfBibitem
\bibitem[not()]{note_moderate}
There are lipids in between the rigid proteins, therefore bending involves
  essentially splaying the lipids and stretching involves essentially changing
  the lipid density. The presence of the proteins should thus not change
  drastically the associated energy penalty per unit area.\relax
\mciteBstWouldAddEndPunctfalse
\mciteSetBstMidEndSepPunct{\mcitedefaultmidpunct}
{}{\mcitedefaultseppunct}\relax
\EndOfBibitem
\bibitem[com()]{compareSeifert}
For $c_0=0$ we recover the dynamical matrix of Ref.~\cite{Seifert93EPL}, except
  that $\eta_s$ includes dilational viscosity.\relax
\mciteBstWouldAddEndPunctfalse
\mciteSetBstMidEndSepPunct{\mcitedefaultmidpunct}
{}{\mcitedefaultseppunct}\relax
\EndOfBibitem
\end{mcitethebibliography}
\bibliographystyle{rsc}

\end{document}